\documentclass[aps,pra,twocolumn,amsmath,amssymb,nofootinbib,superscriptaddress]{revtex4-1}

\newcommand{\ket}[1]{|#1\rangle}

\usepackage[pdftex]{graphicx}
\usepackage{mathtools}
\usepackage{mathrsfs}
\usepackage{hyperref}
\hypersetup{colorlinks = true, linkcolor = [rgb]{0.19411,0.51882,0.667058}, urlcolor = [rgb]{0.125490,0.29542,0.1647058}, citecolor = [rgb]{0.75882,0.37411,0.14117}}
\begin{document}

\bibliographystyle{apsrev}

\title{Efficient recycling strategies for preparing large Fock states from \\ single-photon sources --- Applications to quantum metrology}

\author{Keith R. Motes}
\email{motesk@gmail.com}
\affiliation{Centre for Engineered Quantum Systems, Department of Physics and Astronomy, Macquarie University, Sydney NSW 2113, Australia}

\author{Ryan L. Mann}
\affiliation{Centre for Quantum Computation and Intelligent Systems, Faculty of Engineering \& Information Technology, University of Technology Sydney, NSW 2007, Australia}

\author{Jonathan P. Olson}
\affiliation{Hearne Institute for Theoretical Physics and Department of Physics \& Astronomy, Louisiana State University, Baton Rouge, LA 70803}

\author{Nicholas M. Studer}
\affiliation{Hearne Institute for Theoretical Physics and Department of Physics \& Astronomy, Louisiana State University, Baton Rouge, LA 70803}

\author{E. Annelise Bergeron}
\affiliation{Hearne Institute for Theoretical Physics and Department of Physics \& Astronomy, Louisiana State University, Baton Rouge, LA 70803}

\author{Alexei Gilchrist}
\affiliation{Centre for Engineered Quantum Systems, Department of Physics and Astronomy, Macquarie University, Sydney NSW 2113, Australia}

\author{Jonathan P. Dowling}
\affiliation{Hearne Institute for Theoretical Physics and Department of Physics \& Astronomy, Louisiana State University, Baton Rouge, LA 70803}

\author{Dominic W. Berry}
\affiliation{Department of Physics and Astronomy, Macquarie University, Sydney NSW 2113, Australia}

\author{Peter P. Rohde}
\email[]{dr.rohde@gmail.com}
\homepage{http://www.peterrohde.org}
\affiliation{Centre for Quantum Computation and Intelligent Systems, Faculty of Engineering \& Information Technology, University of Technology Sydney, NSW 2007, Australia}

\date{\today}

\frenchspacing

\begin{abstract}
Fock states are a fundamental resource for many quantum technologies such as quantum metrology. While much progress has been made in single-photon source technologies, preparing Fock states with large photon number remains challenging. We present and analyze a bootstrapped approach for non-deterministically preparing large photon-number Fock states by iteratively fusing smaller Fock states on a beamsplitter. We show that by employing state recycling we are able to exponentially improve the preparation rate over conventional schemes, allowing the efficient preparation of large Fock states. The scheme requires single-photon sources, beamsplitters, number-resolved photo-detectors, fast-feedforward, and an optical quantum memory.
\end{abstract}

\maketitle

\section{Introduction}

Fock states form a ubiquitous resource for many quantum technologies \cite{bib:NielsenChuang00, bib:KokLovett}, ranging from quantum communication and quantum cryptography, to quantum-enhanced metrology \cite{bib:Kapale07, bib:yurke1986input, bib:yuen1986generation, bib:dowling1998correlated, bib:gerry2001generation}, novel interferometric protocols \cite{bib:MagdaLeap}, and all-optical quantum information processing \cite{bib:KLM01}. While much progress has been made in single-photon source technology, preparing large photon-number Fock states remains challenging. Large Fock states can be relatively easily prepared from single-photon Fock states using non-deterministic linear optics, or via heralded spontaneous parametric down-conversion (SPDC). However, scalability is a problem owing to the inefficient (inverse exponential) scaling of the success probability against the desired number of photons.

It is known that quantum enhanced metrology is optimal (i.e. it reaches the Heisenberg limit of phase sensitivity) when NOON states are used \cite{bib:dowling2008quantum}. Creating NOON states with large photon number can be as hard as building a universal optical quantum computer, as it requires many of the same technologies, such as a quantum memory, and feedforward. Nonetheless, one needs Fock states with large photon number to first generate NOON states with large photon number for quantum enhanced metrology \cite{bib:Kapale07}. Thus, efficient schemes for generating Fock states with large photon number, as presented in this manuscript, are an important stepping stone for realizing optimal quantum enhanced metrology.

Here, we present and analyze a bootstrapped protocol for iteratively building up large photon-number Fock states from a resource of single photons. We show that the bootstrapped technique exponentially improves scalability compared to na\"ive single-shot preparation techniques, allowing efficient (polynomial time) state preparation. The experimental requirements are largely the same as for universal linear optics quantum computing (LOQC) \cite{bib:KLM01}, which, while presently challenging, will be viable when all-optical quantum computing eventuates.

This paper is structured as follows. In Sec. \ref{sec:SPDC}, we review the most common form of Fock state preparation --- heralded spontaneous parametric down-conversion --- where we condition upon detecting some number of photons in one mode, guaranteeing the same number of photons in the other. In Sec. \ref{sec:single_shot}, we present a na\"ive brute-force approach to preparing large Fock states via post-selected linear optics --- a technique which is inefficient, requiring exponential resources in both time and resource states. In Sec. \ref{sec:bootstrapped}, we present an improved technique, based on post-selected linear optics, where larger Fock states are progressively built up by combining smaller Fock states. We also introduce a recycling protocol, which exponentially improves the efficiency of state preparation compared to the previously described schemes. In Sec. \ref{sec:strategies}, we discuss fusion, i.e. how does the choice for the order in which to fuse Fock states affect resource scaling? Specifically we describe our fusion protocol, speak about some analytic approximations, discuss specific fusion strategies that we implement, and talk about hybrid schemes, where, instead of beginning with a resource of single photons, we begin with a resource of larger Fock states, thereby reducing the number of operations required to prepare the larger states. In Sec. \ref{sec:reduction}, we discuss the converse protocol, where we wish to reduce photon number, allowing prepared Fock states with more than the desired photon-number to be shrunk to the desired target size. In Sec. \ref{sec:expImper} we provide an analysis of imperfections in our scheme for guidance with experimental implementations. We present all of our simulation results in Sec. \ref{sec:results}, where we numerically simulate the different Fock state preparation algorithms. We conclude in Sec. \ref{sec:conclusion} and discuss issues that require consideration in future experimental implementations, specifically inefficiencies and imperfect photon mode-overlap.

\section{Post-selected spontaneous parametric down-conversion} \label{sec:SPDC}

\begin{figure}[!htb]
\includegraphics[width=0.65\columnwidth]{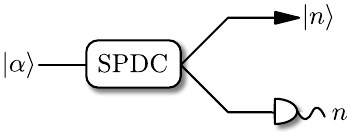}
\caption{Conditional preparation of an $n$-photon Fock state via post-selected SPDC. A non-linear crystal is pumped with a coherent state $\ket\alpha$, yielding a two-mode state given by a correlated superposition in the photon-number basis. Due to the perfect photon-number correlation between the two output modes, detection of $n$ photons in the second output mode guarantees the preparation of $n$ photons in the first. While this works in principle for arbitrary $n$, the success probability decreases exponentially with $n$.} \label{fig:SPDC}
\end{figure}

Perhaps, the most trivial approach to preparing large photon-number Fock states is to employ post-selected, heralded SPDC. Here, we exploit the photon-number correlations between the signal and idler modes of the SPDC output state. That is, upon detecting $n$ photons in one mode, we are guaranteed of having prepared exactly $n$ photons in the other (assuming perfect detection efficiency), as shown in Fig. \ref{fig:SPDC}. This approach has been experimentally demonstrated for small Fock states of up to three photons \cite{bib:MerlinCooper13}.

\begin{figure}[!htb]
\includegraphics[width=0.7\columnwidth]{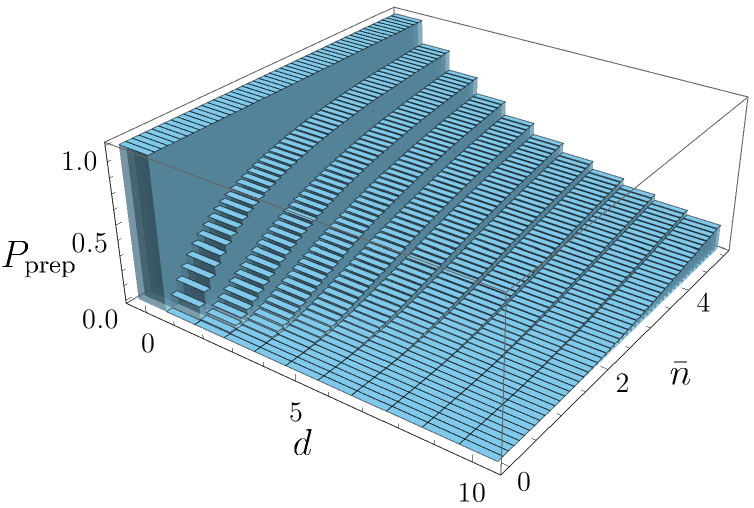}
\caption{The probability of preparing at least $d$ photons in the first mode of an SPDC source with mean photon number $\bar{n}$, by post-selecting upon measuring at least $d$ photons in the other mode.} \label{fig:SPDC_prep}
\end{figure}

An SPDC source coherently down-converts photons from a coherent pump source with amplitude $\alpha$ into photon pairs across the signal and idler output modes via a second-order non-linear process. This evolution is given by a Hamiltonian of the form
\begin{align}
\hat{H}_\mathrm{SPDC} = \chi\hat{a}_\mathrm{pump} \hat{a}_\mathrm{signal}^\dag \hat{a}_\mathrm{idler}^\dag + \mathrm{h.c.},
\end{align}
where $\chi$ is the interaction strength, which depends on the non-linear material. This yields an output state, which is a two-mode superposition in the photon-number basis,
\begin{align}
\ket{\psi_\mathrm{SPDC}} = \sum_{n=0}^\infty \lambda_n \ket{n,n},
\end{align}
where the photon-number distribution is given by
\begin{align}
|\lambda_n|^2 = \frac{1}{\bar{n}+1} \left(\frac{\bar{n}}{\bar{n}+1}\right)^n,
\end{align}
and $\bar{n}$ is the mean photon number, which is a function of the pump-power $\alpha$. Evidently, there is perfect photon-number correlation between the two output modes, providing a direct approach to non-deterministically preparing Fock states of arbitrary photon number. Suppose we wish to prepare large photon-number Fock states, with a target of at least $d$ photons. Then the probability of successful preparation is
\begin{align}
P_\mathrm{prep}(d) = \sum_{n=d}^\infty |\lambda_n|^2 = \left(\frac{\bar{n}}{\bar{n}+1}\right)^d,
\end{align}
which decreases exponentially with $d$, and therefore, the average number of trials required until success scales exponentially with $d$. Furthermore, in present-day laboratories, \mbox{$\bar{n} \ll 1$} \cite{bib:finger15}, making this approach unviable for large $d$. In Fig. \ref{fig:SPDC_prep}, we illustrate the preparation probability for modest values of $d$.

\section{Single-shot preparation via post-selected linear optics} \label{sec:single_shot}

\begin{figure}[!htb]
\includegraphics[width=0.55\columnwidth]{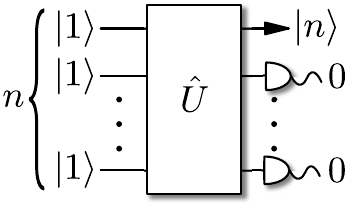}
\caption{Single-shot preparation on an $n$-photon Fock state from $n$ single photons. A photon is incident upon each input mode, and we post-select upon detecting vacuum in all but one of the output modes, where all $n$ photons must have exited the remaining output mode.} \label{fig:single_shot}
\end{figure}

Using only single-photon sources as a resource (either from SPDC or any other source technology), the obvious way in which to prepare large photon-number Fock states using post-selected linear optics, is to simply input single-photon states into each of the input modes of the network, and post-select upon detecting vacuum at all but one of the output modes. When this post-selection succeeds, conservation of photon number ensures that all photons exit through the unmeasured mode. This is illustrated in Fig. \ref{fig:single_shot}.

Let us consider an $n$-mode interferometer with exactly one photon at each input mode. Then the input state is of the form
\begin{align}
\ket{\psi_\mathrm{in}} = \ket{1}^{\otimes n} = \hat{a}^\dag_1 \dots \hat{a}^\dag_n \ket{0}^{\otimes{n}} =\left[\prod_{i=1}^n \hat{a}^\dag_i \right] \ket{0}^{\otimes n},
\end{align}
where $\hat{a}^\dag_i$ is the photonic creation operator for the $i^{th}$ mode.

Next, we apply a linear optics network, implementing the unitary map on the photonic creation operators
\begin{align}
\hat{U} \hat{a}_i^\dag \hat{U}^\dag \mapsto \sum_{j=1}^n U_{i,j}\hat{a}^\dag_j,
\end{align}
yielding
\begin{eqnarray}
\ket{\psi_\mathrm{out}} &=& \hat{U} \ket{\psi_\mathrm{in}} \nonumber \\
&=& \left[ \prod_{i=1}^n \sum_{j=1}^n U_{i,j} \hat{a}^\dag_j \right] \ket{0}^{\otimes n}.
\end{eqnarray}
Post-selecting upon all photons being present in the first mode (i.e. vacuum in all other modes), the projected state is
\begin{eqnarray}
\ket{\psi_\mathrm{proj}} &=& \left[ \prod_{i=1}^n U_{i,1} \hat{a}^\dag_1 \right] \ket{0}^{\otimes{n}} \nonumber \\
&=& \sqrt{n!} \left[\prod_{i=1}^n U_{i,1} \right] \ket{n}\ket{0}^{\otimes{n-1}},
\end{eqnarray}
and the probability of this event occurring is
\begin{align}
P_\mathrm{bunch} = n! \left| \prod_{i=1}^n U_{i,1} \right|^2.
\end{align}
In the case of a balanced interferometer (i.e. one where each input photon has equal amplitude of reaching the first output mode --- which maximizes $P_\mathrm{bunch}$, following from the triangle inequality --- we have
\begin{align}
|U_{i,1}| = \frac{1}{\sqrt{n}} \,\forall\, i,
\end{align}
and then
\begin{align} \label{eq:single_shot_P}
P_\mathrm{bunch} = n! \left| \prod_{i=1}^n \frac{1}{\sqrt{n}} \right|^2 = \frac{n!}{n^n} \sim \frac{\sqrt{n}}{e^n}.
\end{align}

Clearly, the probability of preparing an $n$-photon Fock state decreases exponentially with $n$. Thus, a realistic experimental implementation would be limited to relatively small $n$.

As a simple example, consider two-photon Hong-Ou-Mandel (HOM) \cite{bib:HOM87} interference. In this instance \mbox{$n=2$}, and
\begin{align}
P_\mathrm{bunch} = \frac{2!}{2^2} = \frac{1}{2},
\end{align}
as expected, since the output state of a HOM interferometer is \mbox{$\ket{\psi_\mathrm{out}} = (\ket{2,0}+\ket{0,2})/\sqrt{2}$}.

\section{Bootstrapped preparation via post-selected linear optics} \label{sec:bootstrapped}

\begin{figure}[!htb]
\includegraphics[width=0.8\columnwidth]{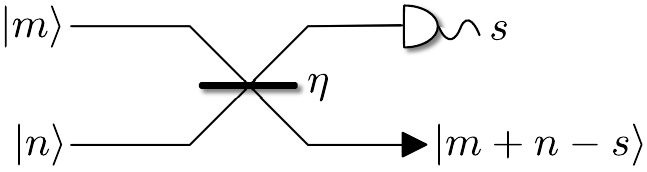}
\caption{The Fock state fusion operation. Two Fock states with $m$ and $n$ photons are mixed on a beamsplitter of reflectivity $\eta$. Upon detecting $s$ photons in the first output mode, an \mbox{$m+n-s$} photon state is prepared in the other mode.} \label{fig:fusion}
\end{figure}

To improve upon the unfavorable scaling of the single-shot approach, we next consider a bootstrapped approach, where progressively larger Fock states are built up using the iterative application of a fusion operation. The fusion operation is simply a beamsplitter, fed with two Fock states with photon numbers $m$ and $n$, where one output mode is measured in the photon-number basis. When no photons are detected, this yields a larger Fock state of size $m+n$. When $s$ photons are detected we have prepared a subtracted state of size \mbox{$m+n-s$}, as shown in Fig. \ref{fig:fusion}.

The input state to the fusion operation is
\begin{eqnarray}
\ket{\psi_\mathrm{in}} &=& \ket{m,n} \nonumber \\
&=& \frac{1}{\sqrt{m!n!}} (\hat{a}^\dag)^m (\hat{b}^\dag)^n \ket{0,0},
\end{eqnarray}
which is incident upon a beamsplitter of reflectivity $\eta$, given by the biased beamsplitter unitary
\begin{align}
U_\mathrm{BS} = \left(\begin{array}{cc}
\eta & \sqrt{1-\eta^2} \\
\sqrt{1-\eta^2} & -\eta 
\end{array}\right),
\end{align}
where local phases are irrelevant. Then the output state is
\begin{align}
\ket{\psi_\mathrm{out}} &= \hat{U}_\mathrm{BS} \ket{\psi_\mathrm{in}} \nonumber \\
&= \frac{1}{\sqrt{m!n!}} (\eta\hat{a}^\dag + \sqrt{1-\eta^2}\hat{b}^\dag)^m \nonumber \\
&\quad\times (\sqrt{1-\eta^2}\hat{a}^\dag - \eta \hat{b}^\dag)^n \ket{0,0} \nonumber \\
&= \frac{1}{\sqrt{m!n!}} \sum_{j=0}^m \binom{m}{j} \eta^j \sqrt{1-\eta^2}^{m-j} (\hat{a}^\dag)^j (\hat{b}^\dag)^{m-j} \nonumber \\
&\quad\times \sum_{k=0}^n \binom{n}{k} \sqrt{1-\eta^2}^k \eta^{n-k} (-1)^{n-k} \nonumber \\
&\quad\times (\hat{a}^\dag)^k (\hat{b}^\dag)^{n-k} \ket{0,0} \nonumber \\
&= \frac{1}{\sqrt{m!n!}} \sum_{j=0}^m \sum_{k=0}^n \binom{m}{j} \binom{n}{k} \eta^{n+j-k} \sqrt{1-\eta^2}^{m+k-j} \nonumber \\
&\quad\times (-1)^{n-k} (\hat{a}^\dag)^{j+k}(\hat{b}^\dag)^{m+n-j-k}\ket{0,0} \nonumber \\
&= \frac{1}{\sqrt{m!n!}} \sum_{j=0}^m \sum_{k=0}^n \binom{m}{j} \binom{n}{k} \eta^{n+j-k} \sqrt{1-\eta^2}^{m+k-j} \nonumber \\
&\quad\times (-1)^{n-k} \sqrt{(j+k)!(m+n-j-k)!} \nonumber \\
&\quad\times \ket{j+k,m+n-j-k}.
\end{align}
We are interested in the case where $s$ photons are measured in the first mode, thereby producing an $s$-photon-subtracted state in the second mode. Thus, we let \mbox{$s=j+k$}, and the unnormalized post-selected state reduces to
\begin{align}
\ket{\psi_\mathrm{ps}}&=\sqrt{\frac{s!(m+n-s)!}{m!n!}} \sum_{j=0}^m \binom{m}{j} \binom{n}{s-j} \eta^{n+2j-s} \nonumber \\
&\quad\times \sqrt{1-\eta^2}^{m+s-2j} (-1)^{n-s+j} \ket{s,m+n-s}. \nonumber \\
\end{align}
The probability of detecting $s$ photons is, therefore,
\begin{align}
P_\mathrm{sub}(s|m,n) &= \eta^{2(n-s)} (1-\eta^2)^{m+s} \frac{s!(m+n-s)!}{m!n!} \nonumber \\
&\quad\times \left| \sum_{j=0}^s \binom{m}{j} \binom{n}{s-j} \left[\frac{\eta^2}{\eta^2-1}\right]^{j} \right|^2.
\end{align}

The conditionally-prepared state will have grown if the $s$-photon-subtracted state is larger than both $m$ and $n$. thus, we require \mbox{$s<m+n-\mathrm{max}(m,n)$}. The probability of preparing a state at least as large as both the input Fock states is
\begin{align} \label{eq:P_grow}
P_\mathrm{grow}(m,n) = \sum_{s=0}^{\mathclap{m+n-\mathrm{max}(m,n)-1}} P_\mathrm{sub}(s|m,n).
\end{align}
In the case of the non-recycled protocol, we only accept the \mbox{$s=0$} outcome, so we eliminate the sum and keep only the \mbox{$s=0$} term.

We wish to maximize this probability such that the likelihood of state growth is maximized, so for each configuration of input Fock states $m$ and $n$, we optimize $\eta$ to maximize $P_\mathrm{grow}$,
\begin{eqnarray}
P_\mathrm{opt}(m,n) &=& \mathrm{max}_\eta [P_\mathrm{grow}(m,n)], \nonumber \\
\eta_\mathrm{opt}(m,n) &=& \mathrm{argmax}_\eta [P_\mathrm{grow}(m,n)].
\end{eqnarray}

\section{Fusion} \label{sec:strategies}

\begin{figure}[!htb]
\includegraphics[width=0.7\columnwidth]{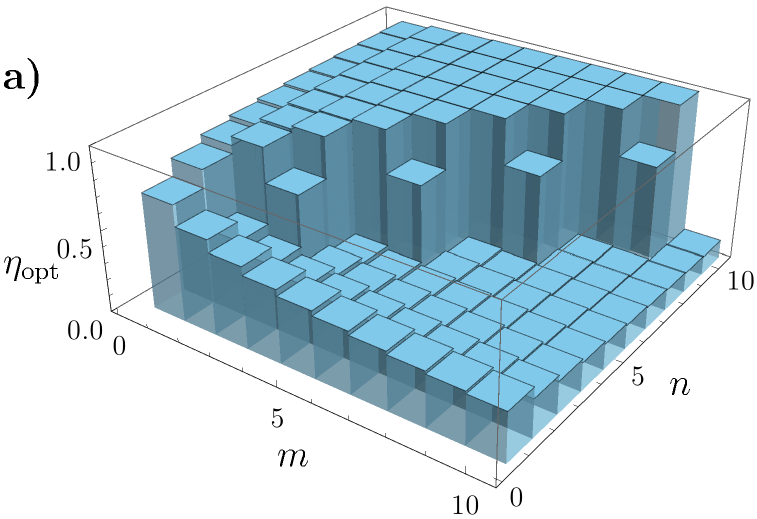} \\
\includegraphics[width=0.7\columnwidth]{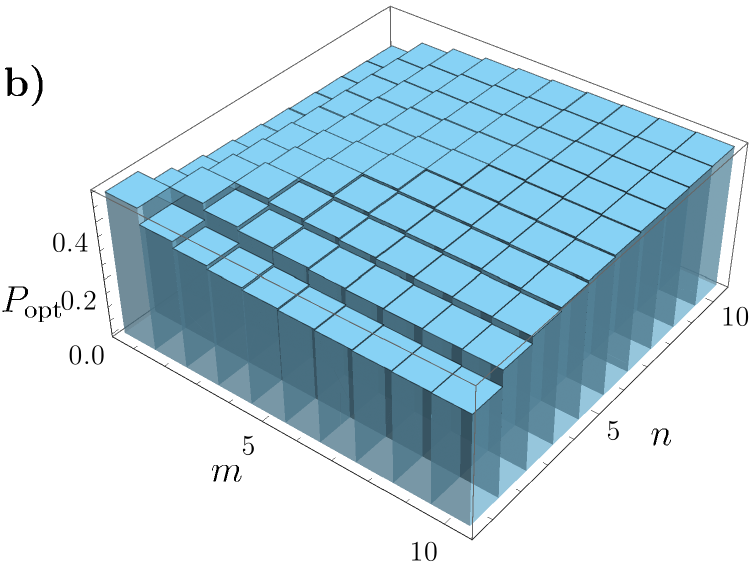} \\
\includegraphics[width=0.7\columnwidth]{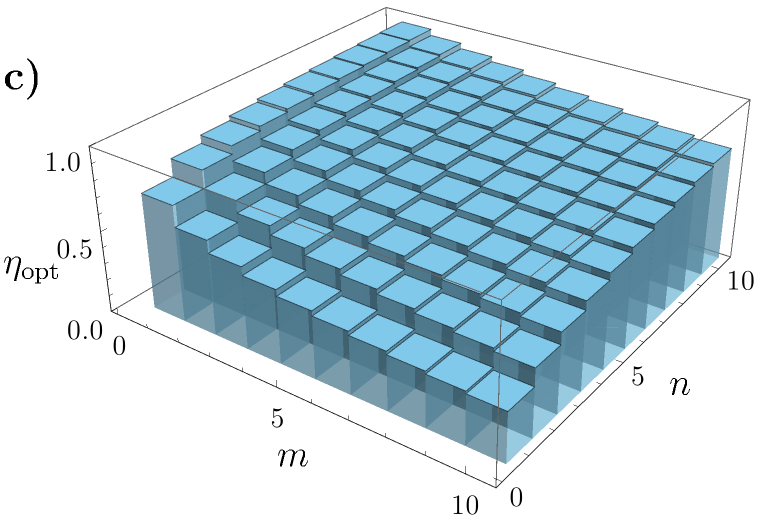} \\
\includegraphics[width=0.7\columnwidth]{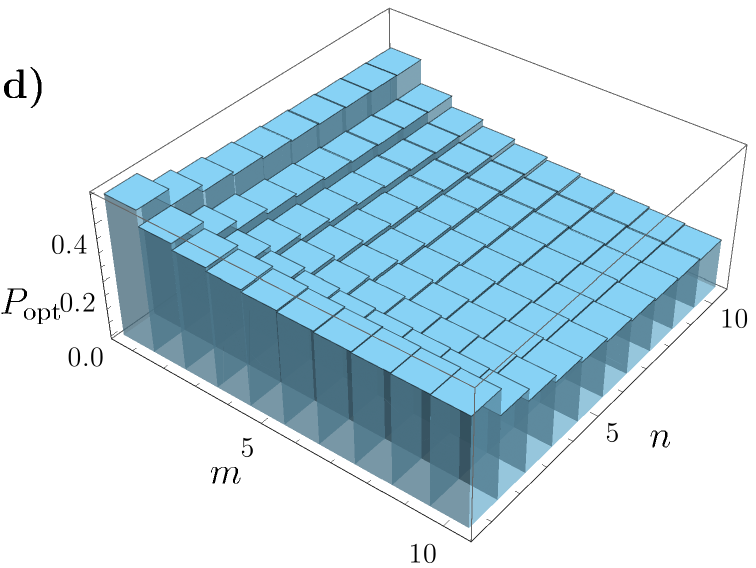}
\caption{(a) The optimal beamsplitter reflectivity ($\eta_\mathrm{opt}$) for maximizing growth probability when fusing two Fock states of photon number $m$ and $n$, accepting all outcomes $s$. (b) The associated growth probability ($P_\mathrm{opt}$) for the fusion operation. (c) The optimal beamsplitter reflectivity for maximizing only the probability of the \mbox{$s=0$} outcome, i.e. for the non-recycled protocol. (d) The associated success probability. Evidently the fusion operation with recycling has higher success probability than without recycling, since we are accepting all events $s$ as opposed to only the \mbox{$s=0$} cases.} \label{fig:opt_eta_P}
\end{figure}

Figure \ref{fig:opt_eta_P} shows the optimized beamsplitter reflectivities and growth probabilities for \mbox{$1\leq m \leq 10$} and \mbox{$1 \leq n \leq 10$}. Evidently, when employing recycling (i.e. accepting all outcomes $s$) the growth probability is maximized when fusing two Fock states of equal photon number, \mbox{$m=n$}. This suggests that the optimal strategy for performing the fusion operations is to always fuse together states of equal size. This is analogous with the cluster state \cite{bib:Raussendorf01, bib:Raussendorf03} literature, where Rohde \& Barrett showed that the balanced strategy \cite{bib:RohdeBarrett07}, where one preferentially bonds cluster states of equal size, is optimal for state growth. In this instance, the only probabilities of interest are \mbox{$P_\mathrm{sub}(s|m,m)$}, and the optimized growth probability is \mbox{$P_\mathrm{opt}(m,m)=1/2\,\,\forall\, m$}. That is, for every fusion operation between two states of equal size, the probability of growing the state is always \mbox{$1/2$}, which is very favorable when dealing with large photon-number states. When not employing recycling (i.e. accepting only the \mbox{$s=0$} outcomes), the growth probability is maximized when we only attempt to fuse a Fock state with the single-photon state (i.e. \mbox{$m=1$} or \mbox{$n=1$}). Rohde \& Barrett refer to this as the modesty strategy, since fusions involve only the smallest states.

Clearly, if we were to progressively build up a large photon-number state by repeatedly utilizing the fusion operation and requiring \mbox{$s=0$}, the success probability of the protocol would decrease exponentially with the number of fusion operations. To improve on this, we borrow the concept of recycling from cluster state protocols \cite{bib:Nielsen04, bib:Gross06, bib:RohdeBarrett07}, where when a fusion operation fails we do not throw away the state and begin from scratch, but rather store the states from failures and reuse them in future fusion operations. Intuitively, we are accepting all detection events, as opposed to only the \mbox{$s=0$} events, which one expects to improve efficiency.

\subsection{Fusion Protocol}
We begin by assuming that we have a resource of single-photon states, that come for free, i.e. these states are trivial to prepare on-demand. Next, we have a series of buckets (quantum memories) containing Fock states of different photon numbers. Let $c_n(t)$ be the number of $n$-photon states that we have in the respective bucket after the $t^{th}$ fusion operation. Since we assume that single-photon states come for free, we let \mbox{$c_1(0)=\infty$}, and initially all other buckets are empty, \mbox{$c_{i>1}(0)=0$}.

We then take two Fock states from the buckets, in accordance with the employed fusion strategy, which we let be $m$ and $n$, and apply the fusion operation between them. With probability \mbox{$P_\mathrm{sub}(s|m,n)$} this prepares the \mbox{$(m+n-s)$}-photon state, which updates the buckets according to
\begin{eqnarray} \label{eq:transitions}
c_m &\to& c_m-1, \nonumber \\
c_n &\to& c_n-1, \nonumber \\
c_{m+n-s} &\to& \left\{ \begin{array}{ll}
    c_{m+n-s} + 1 & \mathrm{with\,\,recycling;} \\
    c_{m+n-s} + \delta_{s,0} & \mathrm{without\,\,recycling}.
 \end{array}
\right.
\end{eqnarray}
That is, we remove a state from each of the $m$- and $n$-photon buckets, and add one more state to the \mbox{$m+n-s$} photon bucket. Thus, the protocol proceeds as a random walk of populations between the buckets. With recycling, we accept all events $s$, and without recycling accept only the $s=0$ outcomes. Obviously the latter case repopulates the larger photon-number buckets with lower probability, reducing the rate of state preparation.

Next, suppose we wish to prepare a resource of Fock states with photon number at least $d$. Then we are interested in the buckets
\begin{align}
c_{\geq d} = \sum_{j=d}^\infty c_j.
\end{align}
The rate at which these states are prepared, per fusion operation, is then
\begin{align}
r(d) = \lim_{t\to\infty} \frac{c_{\geq d}(t)}{t},
\end{align}
where $t$ is the number of fusion operations that have been applied. We consider the \mbox{$t\to\infty$} limit to establish the steady state flow dynamics of states through the buckets under the action of the random walk.

\subsection{Analytic Approximations}
For some simplified schemes, we can establish analytic results that show that the rate is improved exponentially over the single-shot case discussed in Sec. \ref{sec:single_shot}.
First we consider a non-recycled scheme, where we attempt to construct a Fock state with $d$ photons, where $d$ is a power of $2$.
For values of $d$ that are not a power of $2$, we can simply construct a Fock state with a number that is the next largest power of $2$.
The rate will be unchanged, and the scaling will not be significantly affected.

As this is a non-recycled scheme we only retain successes.
We therefore fuse single-photon states until we obtain $2$-photon states, fuse $2$-photon states until we obtain $4$-photon states, and so forth, which is why we consider powers of two.
As was found above (see Fig. \ref{fig:opt_eta_P}), the success probability is maximized for 50/50 beam splitters when using equal photon numbers, so we use 50/50 beam splitters.
To estimate the rate, we will estimate the average number of single-photon states needed to produce one $d$-photon state.
Then the preparation rate of $d$-photon Fock states per \emph{fusion operation} will have scaling equal to the inverse of this number.
That is because there can be no more than a factor of $2$ between the number of single-photon states used and the number of fusion operations.

To show that result, consider first the ideal case where every fusion operation is successful.
Then for $d$ single-photon states, there would be $d/2$ fusion operations, and $d/4$ fusion operations on the 2-photon states, and so forth.
Adding them together for $d$ a power of two gives a total number of fusion operations equal to $d-1$, or one less than the number of single-photon states.
As the success rate is reduced, the number of fusion operations can only be reduced for a given number of single photons.
Hence the number of fusion operations cannot be any larger than the number of single photons.
For this scheme we perform fusion operations on all pairs of single photons, so the number of fusion operations must be at least half the number of single-photon states.

Now we estimate the average number of single-photon states required to produce one $d$-photon state.
The expected number of attempts to fuse two $d/2$-photon states to produce the $d$-photon state will be $1/P_{\rm sub}(0|d/2,d/2)$.
This corresponds to an expected number of $d/2$-photon states of $2/P_{\rm sub}(0|d/2,d/2)$, as there are two in each attempt.
Then, the expected number of $d/4$-photon states required to produce each $d/2$-photon state is $2/P_{\rm sub}(0|d/4,d/4)$.
As a result, the expected number of $d/4$-photon states required to produce one $d$-photon state is $4/[P_{\rm sub}(0|d/2,d/2)P_{\rm sub}(0|d/4,d/4)]$.
Continuing this reasoning, the expected number of single photons required to produce one $d$-photon state is
\begin{align}
\label{expsin}
\prod_{j=1}^{\log_2 d} \frac 2{P_{\rm sub}(0|d/2^{j},d/2^{j})}.
\end{align}

To estimate this expected number of photons, we can use
\begin{equation}
P_{\rm sub}(0|n,n) = 2^{-2n} \frac{(2n)!}{(n!)^2} \sim \frac {1}{\sqrt{\pi n}},
\end{equation} 
where the approximation is via Stirling's formula.
Using this approximation with Eq. (\ref{expsin}) gives the expected number of single-photon states scaling as
\begin{eqnarray}
2^{\log_2 d} \prod_{j=1}^{\log_2 d} \sqrt{\pi d/2^j} &=& d (\pi d)^{(\log_2 d)/2} 2^{-\log_2 d (\log_2 d+1)/4)} \nonumber \\
&=& d^{3/4+(\log_2 \pi)/2+(\log_2 d)/4}.
\end{eqnarray}
Testing this expression numerically, we find that the expected number of single photons is about $1.2777$ times this value.

As discussed above, the preparation rate of $d$-photon Fock states per fusion operation will scale as the inverse of this expression, and is therefore
\begin{equation}
r(d) \propto \frac 1{d^{3/4+(\log_2 \pi)/2} d^{(\log_2 d)/4}} .
\end{equation}
There is an exponential improvement over the case with just a single interferometer and single-shot preparation.
The scaling is not exponential in $d$, but it is also not polynomial in $d$, because the power is logarithmic in $d$.

\begin{figure}[b]
\includegraphics[width=\columnwidth]{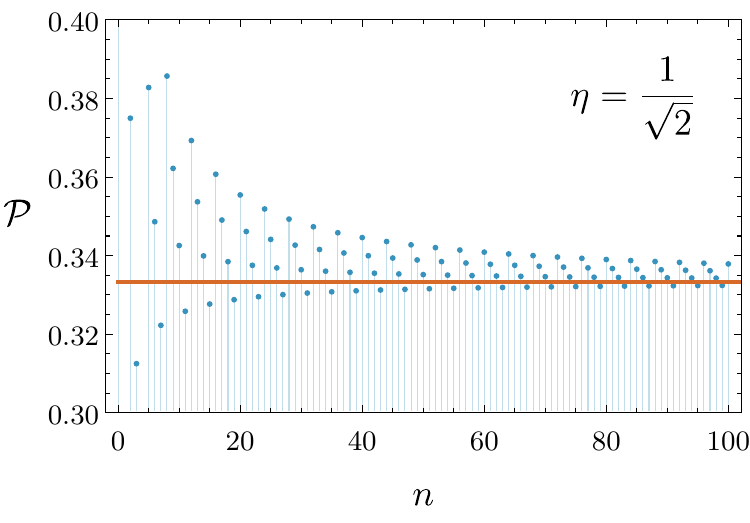}
\caption{The probability of an improved recycling scheme with $\eta=1/\sqrt{2}$, where we require that the number of photons lost is no more than $n/2$ when fusing two $n$-photon states. We see that this probability approaches $\sim1/3$ as $n\to\infty$.} \label{fig:limited_recycling}
\end{figure}

For a further improvement, we can add limited recycling.
Rather than just requiring zero photons to be lost at each stage, we require that the number of photons lost is no more than $n/2$ when fusing two $n$-photon states so that the probability of success is given by 
\begin{equation}
\mathcal{P}=\sum_{s=0}^{\mathrm{Floor}(n/2)}P_\mathrm{sub}(s|n,n),
\end{equation}
with $\eta=1/\sqrt{2}$.
Then we find numerically that the probability of success approaches $\sim1/3$ in the limit $n\to\infty$ as can be seen in Fig. \ref{fig:limited_recycling}. An analytic solution for this result currently eludes us so we have provided numerical evidence. 
It is smaller for smaller values of $n$, but because we are calculating the scaling for large $d$ we will take the probability of success to be $1/3$.
Without loss of generality we can require that on success the photon number is $\lceil 3n/2 \rceil$.
If the photon number is larger than that, we can reduce the photon number with the state reduction scheme described in Sec. \ref{sec:reduction}.
Now, in order to obtain photon number $d$, we need a number of levels scaling as $\log_{3/2} d$, rather than $\log_2 d$.
However, the multiplying factor for the number of repetitions at each stage is smaller.

Therefore, the number of single photons required to obtain a single $d$-photon Fock state scales as
\begin{equation}
6^{\log_{3/2}d} = d ^{\log_{3/2}6} = d^{4.419\ldots}.
\end{equation}
Testing this expression numerically, the number of single photons required is about $47$ times less.
The corresponding rate for this scheme then scales as
\begin{equation} \label{eq:MFscaling}
r(d) \propto d^{-4.419\ldots} .
\end{equation}
This scaling is again an exponential improvement, and is now strictly polynomial.



\subsection{Fusion Strategies}
An analytic bound for more advanced recycling strategies is nontrivial and we instead simulate these protocols as classical Markov processes between the buckets, with transition probabilities given by the parameters $P_\mathrm{sub}(s|m,n)$, and transition rules from Eq. (\ref{eq:transitions}).

While Fig. \ref{fig:opt_eta_P} intuitively hints that the balanced strategy may be optimal, we consider several other strategies as a comparison to provide further insight into the importance of fusion strategy. We will consider four recycling strategies in total:
\begin{itemize}
\item Balanced: Fuse the largest two available states of equal size, \mbox{$m=n$}. This strategy is based on the observation of Fig. \ref{fig:opt_eta_P} that fusing states of equal size maximizes the growth probability, $P_\mathrm{grow}$.
\item Modesty: Always attempt to grow our state by a single photon, by fusing the state $m$ with a single-photon state, \mbox{$n=1$}.
\item Random: Randomly choose any two available states, irrespective of their relative sizes.
\item Frugal: Same as balanced, except that it does not attempt to fuse two equally sized states if \mbox{$m=n>\lfloor d'/2 \rfloor$}, where \mbox{$d' \geq d$}. For states of size \mbox{$n > d'/2$}, it instead attempts to fuse available states such that \mbox{$d\leq m+n \leq d'$}. This is because larger number states are costly to prepare, so it is wasteful to fuse two states with a total photon number well in excess of the target $d$. 
\end{itemize}
The optimization technique for the frugal strategy is different than for the other strategies, which use $P_{\mathrm{grow}}$. If the total input photon number \mbox{$m+n \geq d$}, then for each configuration of input Fock states $m$ and $n$, we optimize $\eta$ to maximize the probability of getting at least $d$ photons. Specifically, we maximize
\begin{align}
	\sum_{s=0}^{\mathclap{m+n-d}}P(s|m,n).
\end{align}

If \mbox{$m+n<d$}, then for each configuration of input Fock states $m$ and $n$, we optimize $\eta$ to maximize the probability of increasing the maximum photon number with increasing weightings for growing the photon number larger. Specifically, we maximize
\begin{align}
	\sum_{s=0}^{\mathclap{m+n-\mathrm{max}(m,n)}}[m+n-s-\mathrm{max}(m,n)]P(s|m,n). \nonumber\\
\end{align}

\subsection{Hybrid schemes} \label{sec:hybrid}

Thus far, we have considered the scenario where a free resource of single-photon states is available. Clearly, this is appropriate when our sources are true single-photon sources. However, emerging photon-source technologies, such as some quantum dot sources, have the ability to prepare small-photon-number Fock states directly. Intuitively, one expects that by starting with a resource of such larger states, by mitigating the need to non-deterministically prepare these states from single photons, we might further improve preparation rates.

This is easily accommodated for in our framework by letting \mbox{$c_x=\infty$} rather than \mbox{$c_1=\infty$}, where $x$ is the number of photons that can be prepared in a single shot. For example, beginning with two-photon sources, we let \mbox{$c_2=\infty$}. Having made this slight adjustment to the protocol, it proceeds as usual using the appropriate fusion strategy.

\section{State reduction} \label{sec:reduction}

In our analysis, we have let $d$ be our target number of photons, and the state preparation rate $r(d)$ is defined to be the rate at which Fock states of size \emph{at least} $d$ are prepared. For some protocols employing large Fock states, this is an appropriate definition. However, for other applications one may require states of size \emph{exactly} equal to $d$. In this instance we would need a protocol by which to reduce the photon number of states larger than $d$ to have photon number exactly $d$.

This may be easily efficiently implemented using post-selected linear optics. We simply impinge the prepared Fock state on an extremely low-reflectivity beamsplitter with vacuum at the other input (\mbox{$n=0$}, \mbox{$\eta\ll 1$}). Because the beamsplitter reflectivity is low, most of the time no photons will be detected in the reflected mode. Sometimes a single photon will be detected. And with higher order probability more than one photon will be detected. By making the reflectivity sufficiently small, these higher order probabilities may be made arbitrarily low, such that with close to unit probability we detect at most one photon. When a single photon is detected we have reduced the photon number by one. We simply apply this protocol repeatedly until the desired number of photons have been subtracted, yielding the Fock state of the desired target photon number. Note that this protocol is efficient, requiring $O(s)$ beamsplitter operations on average when attempting to subtract $s$ photons.

Because the Fock states are only ever mixed with the vacuum state in the state reduction protocol, there are no mode-matching requirements, making state reduction somewhat experimentally easier to implement than state growth.

\section{Experimental Imperfections} \label{sec:expImper}

There are many experimental errors to consider in this scheme, including loss, 
mode-mismatch, and detector and photon source inefficiencies, among others. The accumulation of errors throughout this scheme will cause incorrect binning of Fock states, which will diminish the fidelity of the desired output state. Loss is the major experimental error that plagues this device, which can occur at any stage of the procedure. There are two distinct loss mechanisms in this scheme: firstly, loss will occur while photons are stored in the quantum memory, and secondly, loss will occur due to inefficient photodetectors during each fusion operation. Losses, in general, will produce a number of photons less than what is expected, contributing to a mixed state, which reduces the fidelity of the desired output state. For a given fusion operation in the presence of loss, if we are given $t=m+n$ total photons, then the probability of retaining $i$ photons and losing $t-i$ photons may be modelled by,
\begin{equation}
p_i=\nu^i(1-\nu)^{t-i}\binom{t}{i},
\end{equation}
where $1-\nu$ is the probability of losing a single photon, and so $\nu$ is the efficiency of a single instance of the fusion operation. When $\nu\approx1$ then $p_t\approx 1-(1-\nu)t$. That is, in the limit of small loss, the probability of losing any photons is proportional to the number of photons. This error happens regardless of whether the loss occurs during detection or in the heralded mode, resulting in a mixture of the desired state with a state containing lower photon number. The total probability of loss for the protocol is then proportional to the average number of photons in a given fusion operation multiplied by the number of fusion operations. 

Another advantage of our scheme is that Fock states have no phase and so the scheme is immune to dephasing. 

\section{Results} \label{sec:results}

\begin{figure}[!htb]
\includegraphics[width=\columnwidth]{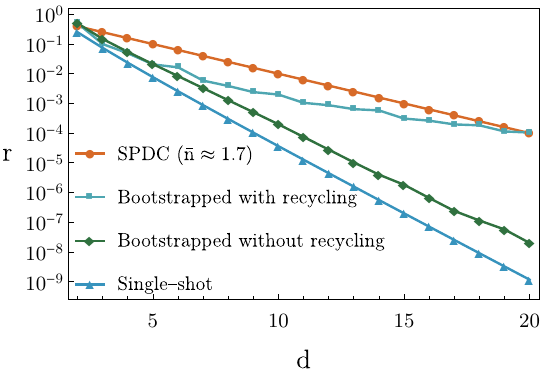}
\caption{Rate of preparation $r$ of Fock states with at least $d$ photons, for the recycled and non-recycled bootstrapped protocols (using the balanced strategy), the single-shot protocol, and the heralded SPDC protocol. In the cases of the two bootstrapped protocols and the SPDC protocol, we observe a strict exponential decrease in the preparation rate against the number of photons. It is evident that state recycling yields a far more favorable scaling than the non-recycled or single-shot protocols. For heralded SPDC, the exponential exhibits a dependence on the mean photon number ($\bar{n}$) of the source, which reflects the SPDC pump-power. Here we have chosen $\bar{n}$ such that the SPDC and recycled bootstrapped balanced protocols have the same 20-photon preparation rate, providing a baseline for the regime of SPDC operation such that it matches the preparation rate of the recycled protocol.} \label{fig:rate}
\end{figure}

In Fig. \ref{fig:rate}, we plot the rate of $d$-photon state preparation, $r(d)$, for both the recycled and non-recycled bootstrapped protocols (Sec. \ref{sec:bootstrapped}), the single-shot protocol (Sec. \ref{sec:single_shot}), and heralded SPDC (Sec. \ref{sec:SPDC}).

In both bootstrapped protocols we employ the balanced fusion strategy and assume an infinite resource of single-photon states, \mbox{$c_1=\infty$}. Here the currency for resource requirements is the number of fusion (i.e. beamsplitter) operations.

In the case of the single-shot protocol, Eq. (\ref{eq:single_shot_P}) gives us the rate of state preparation in units of the number of trials of the entire interferometer. To convert this to the currency of number of beamsplitters, we recognize that the interferometer shown in Fig. \ref{fig:single_shot} can be most easily constructed from $d$ beamsplitters in a linear array to progressively route every photon to the top mode. Thus, the rate of state preparation, as measured by number of beamsplitter operations, is simply given by Eq. (\ref{eq:single_shot_P}) with an additional factor of \mbox{$1/d$}, yielding \mbox{$r(d) = d!/d^{d+1}$}.

In the case of heralded SPDC, there is no direct conversion for resource requirements, since the SPDC protocol employs neither beamsplitters nor single photons as a resource. Rather, the currency is the number of repetitions of the SPDC source (given by the pump repetition rate). Thus, we use the number of repetitions as the resource, bearing in mind that this resource has a different interpretation than the resource usage in the other protocols. In Fig. \ref{fig:rate}, we have chosen the mean photon-number, $\bar{n}$, such that SPDC has the same 20-photon preparation rate as the balanced bootstrapped protocol with recycling, in which case we see that \mbox{$\bar{n}\approx 1.7$} is a threshold, above which SPDC is more efficient than the recycled bootstrapped protocol, and below which is less efficient. This regime for mean photon number is already well beyond what is typically employed in present-day experiments.

It is clear from the log plot that in all cases other than the balanced recycled protocol, the rate of state preparation decreases exponentially with $d$. However, it is evident that recycling substantially improves the preparation rate compared to the non-recycled and single-shot approaches. For example, for 20-photon state preparation, state recycling improves the preparation rate by a factor of \mbox{$\approx 10^5$} over the single-shot protocol.

\begin{figure}[!htb]
\includegraphics[width=\columnwidth]{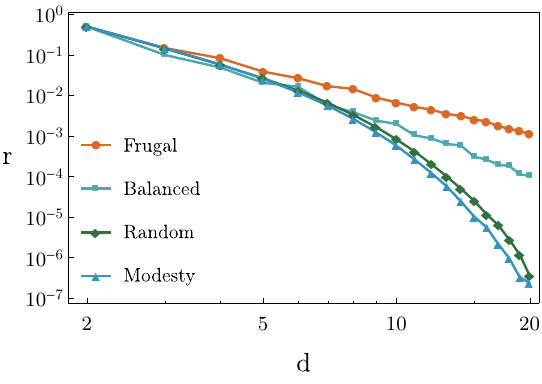}
\caption{Comparison of the strategies when employing state recycling, showing the strategies frugal, balanced, random, and modesty. The linear curves in the log-log plot are indicative of polynomial scaling of the preparation rate against the desired number of photons when employing the frugal and balanced strategies, whereas the random and modesty strategies exhibit exponential scaling.} \label{fig:rate_loglog}
\end{figure}

In Fig. \ref{fig:rate_loglog}, we show the recycled protocol for the various fusion strategies introduced earlier in Sec. \ref{sec:strategies}. From the log log plot, it is evident that the preparation rate exhibits polynomial scaling against $d$ when employing the frugal and balanced recycled strategies, as opposed to the exponential scaling of the random and modesty recycled strategies, and the single-shot, non-recycled and SPDC protocols. This represents an exponential efficiency improvement in the frugal and balanced recycled protocols. The associated scaling for the Frugal and Balanced strategies goes as $\sim1/d^{2.8}$ and $\sim1/d^{3.7}$ respectively, which is significantly better than the scaling in the doubling approach that led to the rate in Eq. (\ref{eq:MFscaling}). 

\begin{figure}[!htb]
\includegraphics[width=\columnwidth]{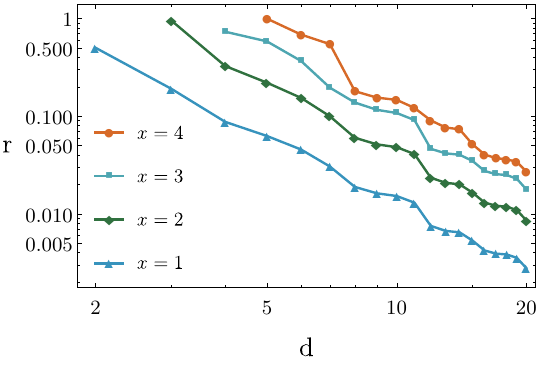}
\caption{Preparation rates for at least $d$ photons for hybrid schemes, where we begin with resource states of different photon numbers $x$ (i.e. \mbox{$c_x=\infty$}), and employ the frugal fusion strategy. Evidently, as we begin with larger starting resource states, the efficiency of the scheme improves, since now the steps that would ordinarily be required to prepare these resource states from single-photon states are mitigated.} \label{fig:hybrid}
\end{figure}

In Fig. \ref{fig:hybrid}, we show the performance of hybrid schemes, where we begin with a resource of Fock states with more than one photon. We only include the results for the recycled frugal protocol, since this was observed to exhibit the best scaling in preparation rates.

\section{Conclusion} \label{sec:conclusion}

We have presented a scheme for preparing large photon-number Fock states non-deterministically from a resource of single-photon states, using post-selected linear optics, by progressively fusing together smaller Fock states into larger ones. We find that by employing state recycling we are able to exponentially improve the state preparation rate --- orders of magnitude for large photon number --- allowing experiments to efficiently (i.e. in polynomial time) prepare much larger Fock states than na\"ive brute-force, single-shot approaches, which require exponential time in general. While some of the experimental requirements, such as quantum memory and fast-feedforward, are presently extremely challenging, these requirements are essentially the same as those for LOQC. Thus, when the tools for optical quantum computing become experimentally viable, all the requirements for this scheme will be in place, and new experiments requiring large photon-number Fock states will be accessible.

In our analysis, we have considered the ideal case, where the resource of single-photon states are perfect $\ket{1}$ states (or more generally $\ket{x}$ states when beginning with $x$-photon states as a resource), and the photo-detectors, beamsplitters and quantum memory have perfect efficiency. In practice, any experimental implementation will exhibit inefficiencies, which would result in output states mixed in the photon-number basis, something which future experimental implementations would need to take into account.

We have also assumed that all photons in the protocol have perfect mode-overlap. In practice, photons will always exhibit some degree of distinguishability, resulting in reduced HOM visibility. This will change the photon-number distribution at the outputs of the fusion operations, affecting the state preparation rate and photon distinguishability of the prepared states. This can be readily modelled using the mode-operator formalism. See Rohde \emph{et al.} \cite{bib:RohdeRalph05, bib:RohdeRalph06, bib:RohdeRalphMunro06, bib:RohdeMauererSilberhorn07} for details on how this technique may be applied to linear optics protocols.

\begin{acknowledgements}
We thank Dave Aldred for helpful discussions. KRM and AG acknowledge the Australian Research Council Centre of Excellence for Engineered Quantum Systems (Project number CE110001013). NMS, EAB, and JPD acknowledge support from the Air Force Office (Grant No. FA9550-13-10098), the Army Research Office (Grant No. W911NF-13-1-0381), the National Science Foundation (Grant No. 1403105), and the Northrop Grumman Corporation. DWB is funded by an ARC Future Fellowship (FT100100761) and an ARC Discovery Project (DP160102426). PPR acknowledges the financial support of Lockheed-Martin.
\end{acknowledgements}

\bibliography{bibliography}

\end{document}